\documentclass[11pt]{article}
\usepackage{amssymb}
\usepackage{amsfonts}
\usepackage{graphicx}
\usepackage{amsmath}

\usepackage{amssymb}

\makeatletter \@addtoreset{equation}{section} \makeatother

\newtheorem{theorem}{Theorem}
\newtheorem{lemma}{Lemma}

\newtheorem{proposition}{Proposition}

\def\P{{\cal P}}

\def\e{\varepsilon}

\begin{document}

\title{Central limit theorem for linear eigenvalue statistics of orthogonally invariant
matrix models}
\author{ M. Shcherbina\\
 Institute for Low Temperature Physics, Kharkov,
Ukraine. \\E-mail: shcherbi@ilt.kharkov.ua
}

\date{}

\maketitle

\begin{abstract}
We prove central limit theorem for linear eigenvalue statistics
 of orthogonally invariant ensembles of random matrices with one interval
 limiting spectrum. We consider ensembles with real analytic potentials
and test functions with  two bounded derivatives.

\end{abstract}

\section{Introduction and main result}\label{sec:1}

In this paper we consider  ensembles of $n\times n$ real symmetric
 matrices $M$  with the probability distribution
\begin{equation}\label{p(M)}
P_{n}(M)dM=Z_{n,\beta}^{-1}\exp \{-\frac{n\beta}{2}\mathrm{Tr}V(M)\}dM,
\end{equation}
where $Z_{n,\beta}$ is the normalization  constant, $V:\mathbb{R}\to \mathbb{R%
}_{+}$ is a H\"{o}lder function satisfying the condition
\begin{equation}\label{condV}
|V(\lambda )|\geq 2(1+\epsilon )\log(1+ |\lambda |).
\end{equation}
and $dM$ means the  Lebesgue measure on the algebraically
independent entries of $M$. In
the case of real symmetric matrices
 $\beta=1$. But since it is interesting to
compare the results with the case Hermitian matrix models, where $\beta=2$,
we keep the parameter $\beta$ in (\ref{p(M)}).

 Let $\{\lambda_i\}_{i=1}^n$ be eigenvalues of $M$.
 Then it is well known (see \cite{Me:91}) that the joint distribution of $\{\lambda_i\}_{i=1}^n$
has the density
\begin{equation}\label{p(la)}
p_{n}(\lambda _{1},...\lambda _{n})=Q_{n,\beta}^{-1}
\exp \{-\frac{n\beta}{2}\sum_{j=1}^{n}V(\lambda _{j})\}\prod_{1\leq j<k\leq
n}|\lambda _{j}-\lambda _{k}|^{\beta},
\end{equation}
where $Q_{n,\beta}$ is the normalizing constant.

 The Normalized  Counting Measure (NCM) of eigenvalues for any
 interval $\Delta \subset \mathbb{R}$ is defined as
\begin{equation}\label{NCM}
{\cal N}_{n}(\Delta )=\sharp \{\lambda _{l}\in \Delta \}/n,
\end{equation}
It is known \cite{BPS,Jo:98} that for any $\beta$
$\quad{\cal N}_{n}(\Delta )$ converges weakly
in probability to a non random measure $ N(\Delta)$, and the limiting
 measure $N$ can be found as a unique minimum of some functional
 on the set of non negative unit measures. The extremum point equation
 for this functional gives us in the case of H\"{o}lder $V'$
\begin{equation}\label{si_eq}
V'(\lambda)=2\int_\sigma\frac{\rho(\mu)d\mu}{\lambda-\mu},\quad \lambda\in\sigma,
\end{equation}
where $\rho$ is the density of $N$ and $\sigma$ is the support of $N$.

For all $\varphi:\mathbb{R}\to\mathbb{R}$ consider a linear statistics
\[N_n[\varphi]=\varphi(\lambda_1)+\dots+\varphi(\lambda_n).\]

It follows from the results of \cite{BPS,Jo:98} that if $V$ is a H\"{o}lder function, then
\[\lim_{n\to\infty}n^{-1}{N}_n[\varphi]=\int\varphi(\lambda)N(d\lambda).\]
Consider the fluctuation of linear eigenvalue statistics
\begin{equation}\label{dot_N}
\dot{N}_n[\varphi]=N_n[\varphi]-E\{N_n[\varphi]\}.\end{equation}

For Hermitian matrix models it was proved by Johansson \cite{Jo:98}
that if $V$ is a real analytic function and the limiting spectrum
$\sigma=[-2,2]$, then for any $\varphi\in C_1[-d-2,2+d]$
$\dot{N}_n[\varphi]$ converges in distribution,
as $n\to\infty$, to a Gaussian random variable.
The limiting variance is the limit, as $n\to\infty$, of
\begin{multline}\label{}
\mathbf{Var}_n[\varphi;V]=E\{\dot{N}_n^2[\varphi]\}=
n(n-1)\int d\lambda_1 d\lambda_2p^{(n)}_{2,\beta}(\lambda_1,d\lambda_2)
\varphi(\lambda_1)\varphi(\lambda_2)\\
+n\int d\lambda_1p^{(n)}_{1,\beta}(\lambda_1)
\varphi^2(\lambda_1)-n^2\bigg(\int d\lambda_1p^{(n)}_{1,\beta}(\lambda_1)
\varphi(\lambda_1)\bigg)^2.
 \end{multline}
 Here and below we denote by $p^{(n)}_{l,\beta}$ the $l$th
marginal density
\begin{equation}\label{marg}
p^{(n)}_{l,\beta}(\lambda_1,...,\lambda_l)=\int d\lambda_{l+1}\dots d\lambda_n
p_n(\lambda_1,\dots,\lambda_n).
\end{equation}

A key role in the proof of CLT and also in the most of  studies
of Hermitian matrix models belongs to the orthogonal polynomials technics,
which allows to write all marginal densities as
\begin{equation}\label{det_f}
p^{(n)}_{l,2}(\lambda_1,...,\lambda_l)=
\frac{(n-l)!}{n!}\det \{K_{n}(\lambda
_{j},\lambda _{k})\}_{j,k=1}^{l},
\end{equation}
where
\begin{equation}\label{K}
K_{n}(\lambda ,\mu;V )=\sum_{l=0}^{n-1}\psi _{l}^{(n)}(\lambda )\psi
_{l}^{(n)}(\mu ).
\end{equation}
is a reproducing kernel of the orthonormal system,
\begin{equation}\label{psi}
\psi _{l}^{(n)}(\lambda )=w_{n}^{1/2}(\lambda )p_{l}^{(n)}(\lambda
),\;\,l=0,...,
\end{equation}
 $p_{l}^{(n)},\;l=0,...$ are orthogonal polynomials on $\mathbb{R}$
associated with the weight $w_{n}(\lambda )=e^{-nV(\lambda )}$
\[
\int p_{l}^{(n)}(\lambda )p_{m}^{(n)}(\lambda )w_{n}(\lambda )d\lambda
=\delta _{l,m}.
\]
 In the Hermitian case it can be proved  that
\begin{multline}\label{var_u}
\frac{d^2}{dt^2}\log E\{e^{t\dot{N}_n[\varphi]}\}=
\mathbf{Var}\{N_{n}[\varphi;V+t\varphi/n]\}\\
=\int d\mu_1 d\mu_2
(\varphi (\mu_1 )-\varphi(\mu_2 ))^2
 {K}_{n}^2(\mu_{1},\mu_{2};V+t\varphi/n).
   \end{multline}
 Hence, to prove CLT we are faced with the problem to study the last integral or to
 prove that $K_n$ does not depend on the "small perturbation" $t\varphi/n$
  in the limit $n\to \infty$. For unitary matrix models it is
 true only in the case (see \cite{Jo:98}), when the  support of $N$ ( limiting NCM)
 consists of one interval. If the
 limiting support consists of two or more intervals, then the r.h.s. of (\ref{var_u})
 has no limit, as $n\to\infty$ (see \cite{P:06}).

In the case of real symmetric matrix  models the situation is more
complicated. According to the result of \cite{Tr-Wi:98},
to study the marginal densities we need to study a matrix kernel of the form
\begin{equation}\label{hat_K}
\widehat K_{n,1}(\lambda,\mu)=\left(\begin{array}{cc}S_n(\lambda,\mu)&
S_nd(\lambda,\mu)\\ -IS_n(\lambda,\mu)& S_n(\mu,\lambda)\end{array}\right),
\end{equation}
where
\begin{equation}\label{S}
S_n(\lambda,\mu)=-\sum_{i,j=0}^{n-1}\psi^{(n)}_i(\lambda)(\mathcal{M}^{(0,n)})^{-1}_{i,j}
(n\e\psi^{(n)}_j)(\mu),
\end{equation}
with
\begin{equation}\label{M}
\mathcal{M}^{(0,n)}=\{M_{j,l}\}_{j,l=0}^{n-1},\quad
M_{j,l}=n(\psi^{(n)}_j,\epsilon\psi^{(n)}_l).
\end{equation}
Here and below we denote
\begin{equation}\label{eps}
\epsilon(\lambda)=\frac{1}{2}\hbox{sign}(\lambda);\quad \epsilon
 f(\lambda)=\int\epsilon(\lambda-\mu)f(\mu)d\mu .
\end{equation}
If we know $\widehat K_{n}(\lambda,\mu)$, then
\begin{equation*}
p^{(n)}_{l,1}(\lambda_1,...,\lambda_l)=
\frac{(n-l)!}{n!}\frac{\partial^l}{\partial\varphi(\lambda_1)\dots\partial\varphi(\lambda_l)}
\hbox{det}^{1/2}\{I+\widehat K_n\widehat\varphi \},
\end{equation*}
where $\widehat\varphi$ is the operator of multiplication by $\varphi$ and
$\widehat K_n:L_2[\mathbb{R}]\oplus L_2[\mathbb{R}]\to L_2[\mathbb{R}]\oplus L_2[\mathbb{R}] $
is an integral operator with the matrix kernel $\widehat K_n(\lambda,\mu)$.

In particular,
\begin{equation}\label{p_1,2}\begin{array}{rcl}
p^{(n)}_{1,1}(\lambda)&=&\dfrac{1}{2n}\hbox{Tr}\widehat K_{n}(\lambda,\lambda),
\\  p_{2,1}^{(n)}(\lambda,\mu)&=&
\dfrac{1}{4n(n-1)}\left[\hbox{Tr}
\widehat K_{n}(\lambda,\lambda)
\hbox{Tr}\widehat K_{n}(\mu,\mu)-2\hbox{Tr}\widehat K_{n}(\lambda,\mu)\widehat K_{n}(\mu,\lambda))\right].
\end{array}\end{equation}
Below  we will use also the following representation of the variance
$\mathbf{Var}\{N_{n}[\varphi _{1};V]\}$
 \begin{proposition} \label{p:var}
\begin{equation}\label{pv.1}
\mathbf{Var}\{N_{n}[\varphi _{1}];V\}
 =\frac{1}{4}%
\int d\mu_1 d\mu_2
(\varphi _{1}(\mu_1 )-\varphi _{1}(\mu_2 ))^2
  \emph{tr\,} \left(\widehat{K}_{n}(\mu_{1},\mu_{2})
   \widehat{K}_{n}(\mu_{2},\mu_{1})\right)
  \end{equation}
\end{proposition}
The structure of the matrix kernel $\widehat K_n$ is studied only
for a few particular ensembles. The case of GOE it was considered in \cite{Tr-Wi:98}.
The case $V(\lambda)=\lambda^{2m}$ for natural $m$ was studied in  \cite{DeG1}.
The case   $V(\lambda)=\frac{1}{4}\lambda^4-\frac{a}{2}\lambda^2$
was studied in \cite{St1}.

Let us set our main conditions.
 \begin{description}
\item[\textbf{ C1.} ] \textit{$V(\lambda )$ satisfies (\ref{condV}) and
is an even analytic function in }
\begin{equation}
\Omega[d,d_1]=\{z:-2-2d\le\Re z\le 2+2d,\,\, |\Im z|\le d_1\}, \quad
d,d_1>0.
\label{Omega}\end{equation}
 \item[\textbf{ C2.}]\textit{  The support $\sigma $ of
IDS of the ensemble consists of a single interval:}
\[
\sigma =[-2,2].
\]
\item[\textbf{C3.}] \textit{ DOS $\rho(\lambda)$ is
strictly positive in the internal points $\lambda\in (-2,2)$ and $\rho(\lambda)\sim
|\lambda\mp 2|^{1/2}$, as $\lambda\sim\pm2$}.
\item[\textbf{C4.}]\textit{ The function
\begin{equation}
u(\lambda)=2\int\log |\mu-\lambda|\rho(\mu)d\mu-V(\lambda)
\label{u}\end{equation}
 achieves its maximum
if and only if $\lambda\in\sigma$. }
\end{description}
It is proved in \cite{APS:01} that these conditions imply that
\begin{equation}\label{rho}
\rho(\lambda)=\frac{1}{\pi}P(\lambda)\sqrt{4-\lambda^2}\mathbf{1}_\sigma,
\end{equation}
where
\begin{equation}\label{P}
P(z)=\frac{1}{2\pi i }\oint_{\mathcal{L} }{\frac{V^{\prime }(z)-V^{\prime
}(\zeta )}{z-\zeta }}\frac{d\zeta }{(\zeta^2-4)^{1/2}}=
\frac{1}{2\pi }\int_{-\pi}^{\pi}{\frac{V^{\prime }(z)-V^{\prime
}(2\cos y )}{z-2\cos y}}dy .
\end{equation}%
Here the contour $\mathcal{L}\subset \Omega[d,d_1]$ and
$\mathcal{L}$ contains inside the interval $(-2,2)$. It is evident that
$P$ is an analytic function in $\Omega[2d/3,2d_1/3]$ and $P(\lambda)\ge\delta>0$,
$\lambda\in\sigma$.

Under these conditions it was proved in \cite{S:07} that
there exists an $n$- independent $C$ such that for even $n$ $||(M^{(0,n)})^{-1}||\le C$ and
\begin{equation}\label{S.1}
S_n(\lambda,\mu)=K_{n}(\lambda,\mu)+r_n(\lambda,\mu)+\tilde r_n(\lambda,\mu),
\end{equation}
where
\begin{equation}\label{r}
r_n(\lambda,\mu)=n\sum_{|k|,|j|\le 2\log^2n} A_{j,k}^{(n)}\psi^{(n)}_{n+j}(\lambda)
\epsilon\psi^{(n)}_{n+k}(\mu),
\end{equation}
\begin{equation}\label{ti_r}
\tilde r_n(\lambda,\mu)=\sum_{j,k=0}^{n-1}\mathcal{E}^{(n)}_{j,k}\psi^{(n)}_{j}(\lambda)
\epsilon\psi^{(n)}_{k}(\mu),\quad ||\mathcal{E}^{(n)}_{j,k}||\le e^{-c\log^2 n}.
\end{equation}
Here and below we denote by $c,C,C_0,C_1,...$ positive $n$-independent constants (different
in  different formulas).

Besides,
\begin{equation}\label{IS.1}
IS_n(\lambda,\mu)=\int\epsilon(\lambda-\lambda')K_{n}(\lambda',\mu)d\lambda'+Ir_n(\lambda,\mu),
+I\tilde r_n(\lambda,\mu),
\end{equation}
where
\begin{equation}\label{Ir}
Ir_n(\lambda,\mu)=\int\epsilon(\lambda-\lambda')r_{n}(\lambda',\mu)d\lambda',\quad
I\tilde r_n(\lambda,\mu)=\int\epsilon(\lambda-\lambda')\tilde r_{n}(\lambda',\mu)d\lambda',
\end{equation}
and
\begin{equation}\label{Sd.1}
S_nd(\lambda,\mu)=-\frac{\partial}{\partial\mu}K_{n}(\lambda,\mu)+
\frac{\partial}{\partial\mu}r_n(\lambda,\mu)+\frac{\partial}{\partial\mu}\tilde r_n(\lambda,\mu) .
\end{equation}

The main result of the present paper is
\begin{theorem}\label{t:CLT}
Consider the orthogonally invariant ensemble of random matrices defined by (\ref{p(M)})-(\ref{p(la)})
with  $V$ satisfying conditions  C1-C4. Then
for any $\varphi\in C_1[-2-\varepsilon,2+\varepsilon]$, growing not faster than polynomial at infinity,
 fluctuations of linear statistics (\ref{dot_N})
converge in distribution, as $n\to\infty$, to a Gaussian
random variable with zero mean and the variance $\mathbf{Var}[\varphi;V]$,
where
\begin{equation}\label{lim_v}
\mathbf{Var}[\varphi;V]=\lim_{n\to\infty}\mathbf{Var}_n[\varphi;V] .
\end{equation}
\end{theorem}

\section{Proof of the main results}\label{sec:2}

\textbf{Proof of Proposition \ref{p:var} .}
 By definition and (\ref{p_1,2}) we have
 \begin{multline}\label{pv.2}
 \mathbf{Var}_n[\varphi;V]=n(n-1)\int d\lambda d\mu\,
 p_{2,1}^{(n)}(\lambda,\mu)\varphi(\lambda)\varphi(\mu)+n\int d\lambda\,
 p_{1,1}^{(n)}(\lambda)\varphi^2(\lambda)\\
 -n^2\int d\lambda d\mu\,
 p_{1,1}^{(n)}(\lambda)p_{1,1}^{(n)}(\mu)\varphi(\lambda)\varphi(\mu)\\=
 -\frac{1}{2}\int d\lambda d\mu\, \hbox{tr\,} \left(\widehat{K}_{n}(\lambda,\mu)
   \widehat{K}_{n}(\mu,\lambda)\right)\varphi(\lambda)\varphi(\mu)+
\frac{1}{2}\int d\lambda\,\hbox{tr\,} \widehat{K}_{n}(\lambda,\lambda)\varphi^2(\lambda)
\end{multline}
But since
\[\int d\mu\, p_{1,1}^{(n)}(\mu)=1,
\quad \int d\mu\, p_{2,1}^{(n)}(\lambda,\mu)=p_{1,1}^{(n)}(\lambda),
\]
we obtain
\[\frac{1}{2}\int\int d\lambda\hbox{tr\,} \widehat{K}_{n}(\lambda,\lambda)=1,\quad
\int d\lambda d\mu\, \hbox{tr\,} \left(\widehat{K}_{n}(\lambda,\mu)
   \widehat{K}_{n}(\mu,\lambda)\right)=\hbox{tr\,} \widehat{K}_{n}(\lambda,\lambda)\]
Using this expression in (\ref{pv.2}) we get (\ref{pv.1}). $\square$

\medskip

The proof of Theorem \ref{t:CLT} is based on the following  lemma
\begin{lemma}\label{l:gen_c1}
Let  for any $\varphi\in C_1[\sigma_d]$, where $\sigma_d=[-d-2,2+d]$
\begin{equation}\label{lg.1}
\mathbf{Var}_n[\varphi;V]\le C\max_{\sigma_d }|\varphi'|^2,
\end{equation}
and for any polynomial $\varphi$ and any $|t|\le A$
\begin{equation}\label{lg.2}
E\{e^{it\dot N_n[\varphi]}\}\to e^{-t^2\mathbf{Var}[\varphi;V]/2},
\end{equation}
Then for any $\varphi\in C_1[\sigma_d ]$ the  limit  in (\ref{lim_v}) exists  and (\ref{lg.2}) is valid.
\end{lemma}
\textbf{Proof.} Since $\varphi\in C_1[\sigma_d ]$, for any $\varepsilon>0$ there exists $\varphi_1$
and $\varphi_2$, such that $\varphi=\varphi_1+\varphi_2$, $\varphi_1$ is a polynomial and $|\varphi_2'|\le \varepsilon$,
it follows from (\ref{lg.1}) and the Schwarz inequality that there exists $C>0$ independent of $\varepsilon$ and $n$
\[
\left|\mathbf{Var}_n[\varphi;V]-\mathbf{Var}_n[\varphi_1;V]\right|\le C\varepsilon
\]
Besides, for any other choice $\tilde\varphi_1$ and $\tilde\varphi_2$ such that
$\varphi=\tilde\varphi_1+\tilde\varphi_2$,
$|\tilde\varphi_2'|\le \varepsilon_1$, we have
\[
\left|\mathbf{Var}_n[\tilde\varphi_1;V]-\mathbf{Var}_n[\varphi_1;V]\right|\le
C(\varepsilon+\varepsilon_1)
\]
Hence, for any choice of polynomials $\{\varphi_n\}_{n=1}^\infty$ such that
$\max|\varphi'-\varphi'_n|\to 0$, as $n\to\infty$, the sequence $\mathbf{Var}_n[\varphi_{1,n};V]$
is fundamental and have a limit independent  of the choice of $\varphi_{1,n}$. This
imply the existence of the limit in (\ref{lim_v}) and that for any $\varphi_1,\varphi_2\in C_1[\sigma_d ]$
\begin{equation}\label{lg.3}
\left|\mathbf{Var}[\varphi_1;V]-\mathbf{Var}[\varphi_2;V]\right|\le C\max_{\sigma_d }
|\varphi_1'-\varphi_2'|
\end{equation}
To prove (\ref{lg.2}) for any
$\varphi$ we fixe any  $\varepsilon>$, choose $\varphi_1$ and $\varphi_2$ like above
and write by the final increments formula and the Schwarz inequality
\[|E\{e^{it\dot N_n[\varphi_1+\varphi_2]}-E\{e^{it\dot N_n[\varphi_1]}\}|\le
|t|E\{\dot N_n[\varphi_2]e^{it\dot N_n[\varphi_1+\xi\varphi_2]}\}\le
A\mathbf{Var}_n^{1/2}[\varphi_2;V]\le CA\varepsilon
\]
Hence, taking the limit $n\to\infty$, we get
\[
e^{-t^2\mathbf{Var}[\varphi_1;V]/2}-CA\varepsilon\le
\liminf_{n\to\infty}E\{e^{it\dot N_n[\varphi]}\}\le
\limsup_{n\to\infty}E\{e^{it\dot N_n[\varphi]}\}\le
e^{-t^2\mathbf{Var}[\varphi_1;V]/2}+CA\varepsilon
\]
Thus, using (\ref{lg.3}) we get (\ref{lg.2}) for any $\varphi\in C_1[\sigma_d ]$. $\square$

 The next lemma will help us to prove (\ref{lg.2}) for polynomial $\varphi$.
\begin{lemma}\label{l:gen_c2}
Let $\{\phi_n(t)\}_{n=1}^\infty$ be a sequence of analytic uniformly bounded functions in the
circle $B_A=\{t:|t|\le A\}$. Assume also that $\phi_n(t)\to\phi(t)$ for any real $t$,
and $\phi(t)$ is also analytic function in $B_A$. Then $\phi_n(t)\to\phi(t)$
for all $t\in B_A$.
 \end{lemma}
\textbf{Proof.} The proof of the lemma is very simple. According to the Arcella theorem, the sequence
 $\{\varphi_n(t)\}$ is weakly compact in $B_A$. But according to the uniqueness
 theorem, the limit of any convergent in $B_A$ subsequence $\{\varphi_{n_k}(t)\}$ must
 coincide with $\varphi(t)$. Hence we obtain the assertion of the lemma.$\square$

\medskip

\noindent\textbf{Proof of Theorem \ref{t:CLT} }
According to the
results of  \cite{APS:01} and \cite{PS:03}, if we restrict the
integration in (\ref{p(la)}) by $|\lambda_i|\le 2+d$, consider
the polynomials $\{p^{(n,d)}_k\}_{k=0}^\infty$ orthogonal on the interval
$\sigma_{d}=[-2-d, 2+d]$ with the weight
$e^{-nV}$ and set $\psi^{(n,d)}_k=e^{-nV/2}p^{(n,d)}_k$,
then for $k\le n(1+\varepsilon)$ with some $\varepsilon>0$
\begin{equation}\label{apr_pol}\begin{array}{l}
\displaystyle \sup_{\lambda\in\sigma_d}|\psi^{(n,d)}_k(\lambda)-
\psi^{(n)}_k(\lambda)|\le e^{-nC},\quad
\sup_{|\lambda|\ge 2+d/2}|\psi^{(n)}_k(\lambda)|\le e^{-nC} .
\end{array}\end{equation}
Hence, if $\mathcal{M}^{(0,n)}_d$  and $S_{n,d}$ are constructed as in (\ref{M}) and (\ref{S})
 for $\sigma_d$, then
\[||\mathcal{M}^{(0,n)}_d-\mathcal{M}^{(0,n)}||\le e^{-nC}, \quad
\max_{\sigma_d}|S_{n,d}(\lambda,\mu)-S_{n,d}(\lambda,\mu)|\le e^{-nC}.
\]
 Therefore  from the very beginning we can take all
integrals in (\ref{p(la)}), (\ref{marg}), (\ref{pv.1}), (\ref{eps}) and (\ref{M})
over the interval $\sigma_d$ and
 then  we can study $\mathcal{M}^{(0,n)}_d$ and $S_{n,d}(\lambda,\mu)$
 instead of $\mathcal{M}^{(0,n)}$ and $S_{n}(\lambda,\mu)$. But
 to simplify notations we omit below the index $d$. Besides, everywhere below
 integrals without limits mean the integrals in $\sigma_d$ and the symbols
 $(.,.)_2$ and $||.||_2$ mean the standard scalar product in $L_2[\sigma_d]$ and the correspondent
 norm.

\medskip

We use  Lemma \ref{l:gen_c2} to prove that for polynomial $\varphi$
\[\phi_n(t)=E\{e^{t\dot N_n[\varphi]}\}\to e^{t^2\mathbf{Var}[\varphi;V]/2},
\quad n\to\infty,\]
where $\mathbf{Var}[\varphi;V]$ is defined in (\ref{lim_v}).

It is evident that
\[|\phi_n(t)|\le |\phi_n(|t|)|+|\phi_n(-|t|)|.\]
Hence to obtain the uniform bound for $\{\phi_n(t)\}_{n=1}^\infty$ for $t\in B_A$
it is enough to find the uniform bound for $\{\phi_n(t)\}_{n=1}^\infty$ with
$t\in[-A,A]$. And to find the last bound and also to prove the convergence
of $\{\phi_n(t)\}_{n=1}^\infty$ for real $t$ it is enough to prove that the sequence
$\{\phi_n''(t)\}_{n=1}^\infty$
 is uniformly bounded for $t\in[-A,A]$ and that
\begin{equation}\label{lim_var}
\lim_{n\to\infty}\phi_n''(t)=\mathbf{Var}[\varphi;V],\quad t\in[-A,A].
\end{equation}
But it is easy to see that
\begin{equation}\label{t.1}
\phi_n''(t)=\mathbf{Var}_n[\varphi;V+t\varphi/n].
\end{equation}
By another words, for our goal it is enough to prove that
under conditions of Theorem \ref{t:CLT}
\begin{equation}\label{t.2}
\lim_{n\to\infty}\mathbf{Var}_n[\varphi;V+t\varphi/n]=\mathbf{Var}_n[\varphi;V] .
\end{equation}
Let us first to transform the expression for $\mathbf{Var}_n[f;V+t\varphi/n]$ given by Proposition \ref{p:var}.
Using (\ref{S.1})-(\ref{Sd.1}) and integrating by parts in terms, containing
$\dfrac{\partial}{\partial\mu}K(\lambda,\mu)$, we get
\begin{multline}\label{t.4}
2\mathbf{Var}_n[f;V+t\varphi/n]=\int d\lambda d\mu\,
S_n(\lambda,\mu)S_n(\mu,\lambda)\Delta_f^2-
\int d\lambda d\mu\,\frac{\partial}{\partial\mu}S_n(\lambda,\mu)
(IS_n(\mu,\lambda)-\epsilon(\mu-\lambda))\Delta_f^2\\
=2\int d\lambda d\mu\, K_n^2(\lambda,\mu)\Delta_f^2+3\int d\lambda d\mu\, K_n(\lambda,\mu)
r_n(\mu,\lambda)\Delta_f^2+\int d\lambda d\mu\, r_n(\lambda,\mu)
r_n(\mu,\lambda)\Delta_f^2\\
-\int d\lambda d\mu\,\frac{\partial}{\partial\mu}r_n(\lambda,\mu)
(IK_n(\mu,\lambda)-\epsilon(\mu-\lambda))\Delta_f^2-
\int d\lambda
d\mu\,\frac{\partial}{\partial\mu}r_n(\lambda,\mu)Ir_n(\mu,\lambda)\Delta_f^2\\
-2\int d\lambda d\mu\,
K_n(\lambda,\mu)(IK_n(\mu,\lambda)-\epsilon(\mu-\lambda))\Delta_f f'(\mu)-
2\int d\lambda d\mu\,
K_n(\lambda,\mu)Ir_n(\mu,\lambda)\Delta_f f'(\mu)\\
+O(\max|f|^2e^{-c\log^2n})=2I_1+3I_2+I_3-I_4-I_5-2I_6-2I_7+O(\max|f|e^{-c\log^2n}),
\end{multline}
where
\begin{equation}\label{Delta}
\Delta_f=f(\lambda)-f(\mu).
\end{equation}
and  $O(\max|f|^2e^{-c\log^2n})$ is a contribution of the terms containing
integrals of $\tilde r_n(\mu,\lambda)$ of (\ref{ti_r}). Note that all integrated
terms here contain $\psi^{(n)}_k(\pm2\pm d)=O(e^{-nc})$ (see
(\ref{apr_pol})). Hence their contribution is $O(e^{-nc})$.

To proceed further let us recall that,
by standard arguments, $\{\psi _{l}^{(n)}\}$ satisfy the recursion formula
\begin{equation}\label{rec}
\lambda\psi _{l}^{(n)}(\lambda )=J^{(n)}_{l}\psi _{l+1}^{(n)}(\lambda )+
q^{(n)}_l\psi _{l}^{(n)}(\lambda )+
J^{(n)}_{l-1}\psi _{l-1}^{(n)}(\lambda ), \quad l=0,1,\dots\quad J^{(n)}_{-1}=0.
\end{equation}
The Jacobi matrix $\mathcal{J}^{(n)}$ defined by this recursion plays an important
role in our proof.
\begin{lemma}\label{l:M}
Consider $\psi^{(n)}_{j}$ and $J^{(n)}_j,q^{(n)}_j$
defined by (\ref{rec}) for the potential $V+t\varphi/n$. Under
conditions of Theorem \ref{t:CLT} there exists $\tilde\varepsilon>0$,
such that for  all $|j|\le \tilde\varepsilon n$
\begin{equation}\label{lim_J}
J^{(n)}_{n+j}=1+\frac{c^{(1)}t+j}{2P(0)n}+r_j^{(1)},\quad
q^{(n)}_{n+j}=\frac{c^{(0)}t}{2P(0)n}+r_j^{(0)},\quad |r_j^{(\alpha)}|\le
C(\frac{j^2}{n^2}+n^{-4/3}),
\quad \alpha=0,1,
\end{equation}
for $|j|\le n^{1/5}$
\begin{equation}\label{d_eps}
\epsilon\psi^{(n)}_{n+j-1}
-\epsilon\psi^{(n)}_{n+j+1}=2n^{-1}\sum_{k>0}{R}_{j-k}\psi^{(n)}_k+n^{-1}\varepsilon_k^{(n)},
\quad ||\varepsilon_k^{(n)}||_2\le n^{-1/9},
\end{equation}
where
\begin{equation}\label{R}
 R_{j}=\frac{1}{2\pi }\int_{-\pi}^{\pi}\frac{e^{ijx}dx}{P(2\cos x)}.
\end{equation}
and the function $P$ is defined in (\ref{P}).
Moreover, there exists $M_{n-j,n-k}^*$ such that for any  $|j|,|k|\le n^{1/5}$
\begin{equation}\label{lim_M}
M_{n-j,n-k}=M_{n-j,n-k}^*+O(n^{-1/9}),\quad M_{n-j,n-k}^*=M_{k-j+1}-\frac{1}{2}(1+(-1)^j)M_{-\infty}
\end{equation}
with
\begin{equation}\label{M_k}
   M_{k}=(1+(-1)^k)\sum_{j=k}^\infty R_{j},\quad
 M_{-\infty}= 2\sum_{j=-\infty}^\infty R_{j},
\end{equation}
\end{lemma}
The proof of the lemma is given in the next section.

 On the basis of the lemma we can prove now  that the last two integrals
 in the r.h.s. of (\ref{t.4}) ($I_6$ and $I_7$) disappear in the
limit $n\to \infty$. Using the Christoffel-Darboux formula it is easy to see that
for this goal it is enough  to prove that for any polynomial $f,g$ and  any
$|j|,|k|\le \log^2 n$
\begin{equation}\label{t.5}\begin{array}{l}
\displaystyle\int d\lambda d\mu\,
\left(\psi^{(n)}_n(\lambda)\psi^{(n)}_{n-1}(\mu)-\psi^{(n)}_n(\mu)\psi^{(n)}_{n-1}(\lambda)\right)
\left(IK_n(\mu,\lambda)-\epsilon(\lambda-\mu)\right)f(\lambda)g(\mu)\to 0\\
\displaystyle n\int d\lambda d\mu\,
\left(\psi^{(n)}_n(\lambda)\psi^{(n)}_{n-1}(\mu)-\psi^{(n)}_n(\mu)\psi^{(n)}_{n-1}(\lambda)\right)
\epsilon\psi^{(n)}_{n+k}(\lambda)\epsilon\psi^{(n)}_{n+j}(\mu)f(\lambda)g(\mu)\to 0
\end{array}\end{equation}
We use  that
\begin{equation}\label{t.6}
IK_n(\mu,\lambda)-\epsilon(\lambda-\mu)=\sum_{k=n}^\infty\epsilon\psi^{(n)}_{k}(\mu)
\psi^{(n)}_k(\lambda)
\end{equation}
in the weak sense. Besides, using the recursion formula (\ref{rec}) we obtain easily that
for polynomial $f$ of the degree $l$

\begin{equation}\label{t.7}
f(\lambda)\psi^{(n)}_{n-\alpha}(\lambda)=\sum_{k=n+\alpha-l}^{j=n+\alpha+l}f_{n-\alpha,j}
\psi^{(n)}_{n-\alpha+j}(\lambda),\quad\alpha=0,1,
\end{equation}
where, according to (\ref{lim_J}), the coefficients $f_{n+\alpha,j}$ have finite limits, as $n\to\infty$.
Using (\ref{t.6}) and (\ref{t.7}) in the first integral of (\ref{t.5}) and integrating
with respect to $\lambda$, we obtain that the first integral is equal to a finite sum
of the terms
\begin{equation}\label{t.8}
\int d\mu\,\epsilon\psi^{(n)}_{n+j}(\mu)\psi^{(n)}_{n-\alpha}(\mu)g(\mu).
\end{equation}
But using the representation of the type (\ref{t.7}) for the polynomial $g$ we obtain
easily that every term of the type (\ref{t.8}) is equal to a finite sum
of the terms
\begin{equation}\label{t.9}
\int d\mu\,\epsilon\psi^{(n)}_{n+j}(\mu)\psi^{(n)}_{n+j'}(\mu)=n^{-1}M_{n+j',n+j} .
\end{equation}
Since by (\ref{lim_M}) $M_{n+j',n+j}$ have finite limits as
$n\to\infty$ we obtain the first line of (\ref{t.5}).

To prove that the second integral in (\ref{t.5}) tends to zero, we also use (\ref{t.7}) and
its analog for $g$. Then we obtain that the second integral is a finite sum with
convergent coefficients of the terms
\[n\int d\lambda d\mu\,\epsilon\psi^{(n)}_{n+k}(\lambda)\psi^{(n)}_{n+k'}(\lambda)
\epsilon\psi^{(n)}_{n+j}(\mu)\psi^{(n)}_{n+j'}(\mu)=n^{-1}M_{n+k',n+k}M_{n+j',n+j}.\]
Similarly to the above we conclude that all these terms tend to zero and so the second
integral in (\ref{t.5}) tends to zero.

\begin{lemma}\label{l:A}
Consider the coefficients $A^{(n)}_{j,k}$ from (\ref{r}) defined for the potential $V+t\varphi/n$.
Under conditions of Theorem \ref{t:CLT} for any  $|j|,|k|\le \log^2 n$ there exists
$A_{j,k}$ independent of $t$ and such that
\begin{equation}\label{lim_A}
|A^{(n)}_{j,k}-A_{j,k}|\le Cn^{-1/9}.
\end{equation}
Moreover, there exists an $n$-independent $c,C$ such that
\begin{equation}\label{b_A}
|A_{j,k}|\le Ce^{-c(|j|+|k|)} .
\end{equation}
\end{lemma}
We prove this lemma in the next section.
\smallskip

According to the above arguments it is clear now that to prove Theorem \ref{t:CLT} it is enough
to prove that for any polynomial $f$ there exist limits for all integral $I_\alpha$,
($\alpha=1,\dots,5$)
from (\ref{t.4}). The existence of the limit of $I_1$ follows from the result of \cite{Jo:98}.
Using representation (\ref{r}) and the Christoffel-Darboux
formula it is easy to understand that  $I_2$ can be represented as a sum of the terms
\begin{equation}\label{t.10}
T_{j,k}:=n\int d\lambda d\mu\,
\left(\psi^{(n)}_n(\lambda)\psi^{(n)}_{n-1}(\mu)\right.
\left.-\psi^{(n)}_n(\mu)\psi^{(n)}_{n-1}(\lambda)\right)
\psi^{(n)}_{n-j}(\lambda)\epsilon\psi^{(n)}_{n+k}(\mu)\frac{\Delta_f^2}{\lambda-\mu}.
\end{equation}
It is evident that if $f$ is a polynomial of the $l$th degree, then
\[\frac{\Delta_f^2}{\lambda-\mu}=\sum_{|p|,|q|\le 2l-1}\tilde f_p(\lambda)\tilde g_q(\mu),\]
where $\tilde f_p$ and $\tilde g_q$ are some fixed polynomial of the degree less than
$2l$. Since we have the bound (\ref{b_A}), it is enough to prove that the limit exists
for any fixed $j,k$, as
$n\to\infty$. But using for (\ref{t.7}) for $\tilde f_p$ and $\tilde g_q$ and
integrating with respect to
$\lambda$, we reduce the existence of the limit of $T_2(j,k)$ to the existence
of the limits of $M_{n-j',n+k}$ for any fixed $j',k$, which follows from Lemma \ref{l:M}.

The existence of the limits of $I_3$ and $I_5$ can be obtained by the same way. To find the limit
of $I_4$ we use first the relation (\ref{t.6}), then (\ref{t.7}) for $f$ and observe that after
integration with respect to $\lambda$ only the finite number of $k$ in the r.h.s. of (\ref{t.6})
give us nonzero contribution. Hence, as above, we   reduce the problem to the existence of
the limits $M_{n-j,n+k}$, which follows from Lemma \ref{l:M}.

To complete the proof of the theorem we are left to prove the estimate (\ref{lg.1}).
It is clear that for this goal it is enough to prove similar estimates for all terms
$I_\alpha$ $\alpha=1,\dots 7$ in (\ref{t.4}). For $I_1$ we have by the Christoffel-Darboux
formula
\[\int d\lambda d\mu\, K_n^2(\lambda,\mu)\Delta_f^2\le
\max_{\lambda\in\sigma_d}|f'|^2\int d\lambda d\mu\, K_n^2(\lambda,\mu)(\lambda-\mu)^2=
2(J^{(n)}_n)^2\max_{\lambda\in\sigma_d}|f'|^2.\]
To prove the estimates for  others $I_\alpha$ let us prove first the following auxiliary
statement
\begin{proposition}\label{p:2}
For any $g$ with $g'$
 bounded in $\sigma_d$ and any  $|j|,|k|\le 2\log^2n$
\begin{equation}\label{t.11}
\bigg|n\int d\mu\, g(\mu)\psi^{(n)}_{n+j}(\mu)\epsilon\psi^{(n)}_{n+k}(\mu)\bigg|\le
C(\max_{\sigma_d }|g'|+\max_{\sigma_d }|g|).
\end{equation}
\end{proposition}
\textbf{Proof of Proposition \ref{p:2}.}
We start from  a simple relation, which follows from the definition of the
operator $\epsilon$ (see \ref{eps}). For any integrable $f,g$
\begin{equation}\label{(ef,eg)}
\int d\lambda\epsilon f(\lambda)\epsilon g(\lambda)=
\frac{1}{4}(\mathbf{1}_{\sigma_d },f)_2(\mathbf{1}_{\sigma_d },g)_2-
\frac{1}{2}\int_{\sigma_d } d\lambda d\mu\, |\lambda-\mu|f(\lambda)g(\mu) .
\end{equation}
In particular, using a simple observation that
$\frac{1}{2}|\lambda-\mu|=(\lambda-\mu)\epsilon(\lambda-\mu)$
and then the definition (\ref{M}), we get
\begin{multline}\label{(e,e)}
\int d\lambda\epsilon \psi^{(n)}_{j}(\lambda)\epsilon \psi^{(n)}_{k}(\lambda)=
\frac{1}{4}(\mathbf{1}_{\sigma_d },\psi^{(n)}_{j})_2
(\mathbf{1}_{\sigma_d },\psi^{(n)}_{k})_2\\-\frac{1}{n}
\left(J^{(n)}_{j}M_{j+1,k}+J^{(n)}_{j-1}M_{j-1,k}-
J^{(n)}_{k}M_{j,k+1}-J^{(n)}_{k-1}M_{j,k-1}\right).
\end{multline}
Since for odd $k$ $(\mathbf{1}_{\sigma_d },\psi^{(n)}_{j})_2=0$, this relation and
(\ref{lim_M}) gives us immediately that for odd $|k|\le n^{1/5}$
\begin{equation}\label{|e_k|}
\int d\lambda(\epsilon \psi^{(n)}_{n+k}(\lambda))^2\le \frac{C}{n}.
\end{equation}
For even $k$ the same relation  can be obtained if we apply the analog of
 (\ref{(e,e)}) to $f(\lambda)=\lambda\psi^{(n)}_{n+k}(\lambda)=
 J^{(n)}_{n+k}\psi^{(n)}_{n+k+1}(\lambda)+J^{(n)}_{n+k-1}\psi^{(n)}_{n+k-1}(\lambda)$ and
 then use (\ref{d_eps}). Remark also that since   (\ref{apr_pol}) yield
\[|\epsilon\psi^{(n)}_{n+k}(2+\lambda)-\epsilon\psi^{(n)}_{n+k}(2+d/2)|\le e^{-nc},
\quad d/2\le\lambda\le d,
\]
by (\ref{|e_k|}), we have
\begin{equation}\label{|e_k(2)|}
n(\epsilon\psi^{(n)}_{n+k}(2+d))^2d/2\le n\int d\mu\,
(\epsilon\psi^{(n)}_{n+k}(\mu))^2+o(1)\le C.\end{equation}
The last bound and (\ref{|e_k|}) imply
one more useful estimate, valid for any
$f$ with bounded derivative
\begin{equation}\label{|e_kf|}
\int d\lambda\left(\epsilon(f \psi^{(n)}_{n+k})(\lambda)\right)^2\le \frac{C}{n}
(\max_{\sigma_d}|f|+\max_{\sigma_d}|f'|)^2.
\end{equation}
Indeed, using that $\psi^{(n)}_{n+k}=(\epsilon\psi^{(n)}_{n+k})'$ and integrating by parts,
it is easy to obtain
\[\epsilon(f \psi^{(n)}_{n+k})=f(\lambda)\epsilon\psi^{(n)}_{n+k}
-\frac{1}{2}f(2+d)\psi^{(n)}_{n+k}(2+d)-\frac{1}{2}f(-2-d)\psi^{(n)}_{n+k}(-2-d)-
\epsilon\left(f'\epsilon\psi^{(n)}_{n+k}\right).\]
Now, taking the square of the r.h.s. and using (\ref{|e_k(2)|}) and (\ref{|e_k|}),
we obtain (\ref{|e_kf|}).

To prove Proposition \ref{p:2} we consider 3 cases:

(a) $j-k$ is  even;

(b) $k$ is even and $j$ is odd;

(c) $k$ is odd and $j$ is even.

\smallskip

 (a) Using (\ref{d_eps}), it is easy to get that
\[
\bigg|n\int d\mu\, g(\mu)\psi^{(n)}_{n+j}(\mu)\epsilon\psi^{(n)}_{n+k}(\mu)-
n\int d\mu\, g(\mu)\psi^{(n)}_{n+k}(\mu)\epsilon\psi^{(n)}_{n+k}(\mu)\bigg|\le
C|k-j|\max_{\sigma_d }|g(\lambda)|.\]
Then, integrating by parts in the second  integral we obtain
\[
n\int d\mu\, g(\mu)\psi^{(n)}_{n+k}(\mu)\epsilon\psi^{(n)}_{n+k}(\mu)=\frac{n}{2}
g(\mu)(\epsilon\psi^{(n)}_{n+k}(\mu))^2\bigg|_{-2-d}^{2+d}-\frac{n}{2}
\int d\mu\, g'(\mu)(\epsilon\psi^{(n)}_{n+k}(\mu))^2.
\]
Relation (\ref{t.11}) follows now from (\ref{|e_k(2)|}) and (\ref{|e_k|}).

(b) Since for even $k$ $\epsilon\psi^{(n)}_{n+k}(0)=0$, using the result
of \cite{De:99} on the asymptotic of orthogonal polynomials,
it is easy to get that for  any $|\mu|\le1$
\[|\epsilon\psi^{(n)}_{n+k}(\mu)|=\bigg|\int_0^{\mu}\psi^{(n)}_{n+k}(\lambda) d\lambda\bigg|\le
\frac{C}{n}.
\]
Hence, if we define
\[\tilde g(\mu)=g(\mu)\mu^{-1}\mathbf{1}_{|\mu|>1}+\frac{1}{2}\left[g(1)(1+\mu)+g(-1)(1-\mu)\right]
\mathbf{1}_{|\mu|\le 1},
\]
so that $g(\mu)=\tilde g(\mu)\mu$ for $|\mu|\ge 1$, then
\begin{equation}\label{t.12}
n\bigg|\int d\mu\, g(\mu)\psi^{(n)}_{n+j}(\mu)\epsilon\psi^{(n)}_{n+k}(\mu)-\int
d\mu\,\mu\tilde g(\mu)\psi^{(n)}_{n+j}(\mu)\epsilon\psi^{(n)}_{n+k}(\mu)\bigg|\le
C\max_{\sigma_d }|g|.
\end{equation}
It is evident that
 $|\tilde g'(\mu)|\le |g'(\mu)|+|g(\mu)|$. Thus, using the
recursion relations (\ref{rec}), we replace the last integral by
\[
n\int
d\mu\, \tilde g(\mu)\left(J^{(n)}_{n+j}\psi^{(n)}_{n+j+1}(\mu)+
J^{(n)}_{n+j-1}\psi^{(n)}_{n+j-1}(\mu))\right)
\epsilon\psi^{(n)}_{n+k}(\mu)d\mu.
\]
 Hence, we obtain again the case (a).

(c) Integrating by parts, we get
\begin{multline*}
n\int d\mu\, g(\mu)\psi^{(n)}_{n+j}(\mu)\epsilon\psi^{(n)}_{n+k}(\mu)=
ng(\mu)\epsilon\psi^{(n)}_{n+k}(\mu)\epsilon\psi^{(n)}_{n+j}(\mu)\bigg|_{-2-d}^{2+d}\\
-n\int d\mu\, g'(\mu)\epsilon\psi^{(n)}_{n+j}(\mu)\epsilon\psi^{(n)}_{n+k}(\mu)
-n\int d\mu\, g(\mu)\epsilon\psi^{(n)}_{n+j}(\mu)\psi^{(n)}_{n+k}(\mu) .
\end{multline*}
The bounds for first two terms in the r.h.s. were found before, and the last integral
corresponds to the case (b).
Thus we have proved (\ref{t.11}).$\square$

\medskip

To find the bound for $I_2$ in (\ref{t.4}) we use the Christoffel-Darboux formula.
Then we are faced with a problem to find the bounds for the terms $T_{j,k}$
of (\ref{t.10}).
But since the function $\Delta_f^2(\lambda-\mu)^{-1}$ for any $\lambda$ has a
derivative, bounded uniformly   with respect to $\lambda,\mu$, we can apply the bound
(\ref{t.11}) for any fixed $\lambda$. We get
\[T_{j,k}\le C\max_{\sigma_d}|f'|^2\int d\lambda|\psi^{(n)}_n(\lambda)|
|\psi^{(n)}_{n-k}(\lambda)|\le C\max_{\sigma_d}|f'|^2,\]
where the last bound is valid  because of the Schwarz inequality.

 The estimates for $I_3$ and $I_5$ follow directly from (\ref{t.11}) and (\ref{b_A}).
For $I_6$ we use the Christoffel-Darboux formula and then the Schwarz inequality.
Then we get
\begin{equation*}
|I_6|^2\le C\max_{\sigma_d}|f'(\mu)|^4\cdot
\bigg(\int  d\mu\,\sum_{k=0}^{n-1}(\epsilon\psi^{(n)}_{n-1}(\mu))^2+C\bigg) .
\end{equation*}
Here the sum with respect to $k$ appears because of integration with respect to $\lambda$
of $IK^2(\mu,\lambda)$ and  $C$ appears because of integration of
$\epsilon^2(\mu-\lambda)$.
But from (\ref{(e,e)}) it is easy to see that
\[\int  d\mu\,\sum_{k=0}^{n-1}(\epsilon\psi^{(n)}_{k}(\mu))^2=
\frac{1}{4}\sum_{k=0}^{n-1}(\mathbf{1}_{\sigma_d },\psi^{(n)}_{k})^2-
\int  d\lambda d\mu\, K_n(\lambda,\mu)(\lambda-\mu)\epsilon(\lambda-\mu).\]
It follows from the Bessel inequality that the sum in the r.h.s. is bounded by
$(\mathbf{1}_{\sigma_d },\mathbf{1}_{\sigma_d })$. In the second integral we
apply the Christoffel-Darboux formula and then (\ref{lim_M}).

For $I_7$ we apply Christoffel-Darboux formula and then the Schwarz inequality.
We obtain
\begin{multline} \label{t.13}
|I_7|\le nC\max_{\sigma_d }|f'|^2
\left(\sum_{j,k,j',k'}A_{j,k}A_{j',k'}\int d\lambda d\mu\,
 \epsilon\psi^{(n)}_{n+j}(\lambda)\epsilon\psi^{(n)}_{n+k'}(\lambda)
 \epsilon\psi^{(n)}_{n+k}(\mu)\epsilon\psi^{(n)}_{n+k'}(\mu)\right)^{1/2}\\
 \le\max_{\sigma_d }|f'|^2,
 \end{multline}
where the last inequality follows from (\ref{|e_k|}).

Now we are left to prove the bound for $I_4$ (see (\ref{t.4})). Remark, that
because of (\ref{apr_pol}) and (\ref{hat_K})-(\ref{p_1,2}) the integrals
in $[2+d/2,2+d]$ and from $[-2-d,-2-d/2]$ in (\ref{t.4}) give us
 $O(e^{-nc})$ terms. Hence, without loss of generality we can replace
 the function  $f$ in these intervals by a linear one in order
  to have   a new function being continuous with a bounded
 derivative and such that $f(2+d)=f(-2-d)=0$.
Then, integrating by parts with respect to $\mu$, we need to control only the
terms, which do not contain $f(\mu)$. But for odd $k$
$\epsilon\psi^{(n)}_{k}(\pm2\pm d)=0$, and if $j$ and $k$ are even, then
$\epsilon\psi^{(n)}_{k}(\mu)\epsilon\psi^{(n)}_{j}(\mu)$ is an even function
and so $\epsilon\psi^{(n)}_{k}(\mu)\epsilon\psi^{(n)}_{j}(\mu)\big|_{-2-d}^{2+d}=0$.
Hence, integrating by parts in $I_4$, we obtain that all integrated terms disappear.
Thus,
\[I_4=-I_2+2\int d\lambda d\mu\, r_n(\lambda,\mu)(IK_n(\mu,\lambda)-\epsilon(\mu-\lambda))
f'(\mu)\Delta_f=-I_2+2I_{4,1}.\]
The bound for $I_2$ was  found before. Hence, we need to find the bound for
$I_{4,1}$. From definitions (\ref{M})  it is evident that $M_{j,k}=-M_{k,j}$
and therefore  from (\ref{S}) we derive
\[ IS_n(\lambda,\mu)=-IS_n(\mu,\lambda)\Leftrightarrow
IK_n(\mu,\lambda)=-IK_n(\lambda,\mu)-Ir_n(\lambda,\mu)-Ir_n(\mu,\lambda).\]
Now if we replace $IK_n(\mu,\lambda)$ by the above expression, then the terms
containing $Ir_n(\lambda,\mu)$ and $Ir_n(\mu,\lambda)$ can be easily estimated
by using (\ref{t.11}) and (\ref{b_A}). Hence we are left to prove the bound for
\begin{multline*}
\bigg|\int d\lambda d\mu\, r_n(\lambda,\mu)IK_n(\mu,\lambda)\tilde f(\lambda)\tilde g(\mu)\bigg|
=n\bigg|\sum_{j,k}A_{j,k}\sum_{l=0}^{n-1}(\tilde
f\psi^{(n)}_{n-j},\epsilon\psi^{(n)}_{l})(\tilde
g\epsilon\psi^{(n)}_{n+k},\psi^{(n)}_{l})\bigg|\\
\le
n\sum_{j,k}|A_{j,k}|\cdot||\epsilon(\tilde f\psi^{(n)}_{n-j}) ||_2||\tilde g\epsilon\psi^{(n)}_{n+k}||_2
\le C (\max_{\sigma_d}|\tilde f|+\max_{\sigma_d}|\tilde f'|)\cdot\max_{\sigma_d}|\tilde g|,
\end{multline*}
where the last bound follows from (\ref{|e_k|}), (\ref{|e_kf|} and (\ref{lim_A})-(\ref{b_A}).
The term with $\epsilon(\lambda-\mu)$ can be estimated similarly.
This completes the proof of Theorem \ref{t:CLT}.

\section{Auxiliary results}
\textbf{Proof of Lemma \ref{l:M}}.
It is proved in \cite{S:07}, that  for $t=0$, representation (\ref{lim_J}) implies
(\ref{d_eps}) and (\ref{lim_M}). If we know
(\ref{lim_J}) for $t\not=0$, then the proofs of
(\ref{d_eps}) and (\ref{lim_M}) coincides with that of \cite{S:07}.
Hence we need only to prove (\ref{lim_J}).

The  idea  is to use the
perturbation expansion of the string equations:
\begin{equation}\begin{array}{l}
V^{\prime }_t({\cal J}^{(n)})_{k,k}=0,\\
\displaystyle J_{k}^{(n)}V^{\prime }_t({\cal J}^{(n)})_{k,k+1} =\frac{k+1}{n}.
\end{array}\label{string}
\end{equation}
Here and below in the proof of Lemma \ref{l:M} we denote $V_t=V+t\varphi$
and by ${\mathcal{J}}^{(n)}$  a semi-infinite Jacobi matrix,
 defined in (\ref{rec}).
Relations (\ref{string})  can be easily obtained from the identity
\[\begin{array}{l}
\displaystyle\int \left( e^{-nV_t(\lambda )}(P_{k}^{(n)}(\lambda
 ))^2\right) ^{\prime }d\lambda =0,\\
\displaystyle\int \left( e^{-nV_t(\lambda )}P_{k+1}^{(n)}(\lambda
)P_{k}^{(n)}(\lambda )\right) ^{\prime }d\lambda =0.
\end{array}\]
We consider (\ref{string}) as a system of nonlinear equations
with respect to the coefficients $J_{k}^{(n)},q_{k}^{(n)}$.
To have zero order expression for $J^{(n)}_{n+k}$
we use the following lemma, proven in \cite{S:05}:

\begin{lemma}\label{l:J=1}
Under conditions C1-C3 for small enough $\tilde {\varepsilon }$ uniformly in
$k:|k|\le \tilde{\varepsilon }n$
\begin{equation} \label{as_J}
\left|q^{(n)}_{n+k}\right|, \left|J^{(n)}_{n+k}-1\right|\le
C\left(n^{-1/4}\log^{1/2} n+(|k|/n)^{1/2}\right).
\end{equation}
\end{lemma}
Denote ${\cal J}^{(0)}$ an infinite Jacobi matrix with constant coefficients
\begin{equation}\label{J^0}
  {\mathcal{J}}^{(0)}_{k,k+1}= {\mathcal{J}}^{(0)}_{k+1,k}=1,
  \quad{\mathcal{J}}^{(0)}_{k,k}=0
\end{equation}
and for any positive $n^{1/3}<<N<n$ define   an infinite Jacobi matrix
$\tilde {\mathcal{J}}(N)$ with the entries
\begin{equation}\label{ti-J}
 \tilde J_k=
  \begin{cases}
   J^{(n)}_{n+k}-1,  & |k|< N , \\
    0, & \text{otherwise}.
  \end{cases}\quad
  \tilde q_k=
  \begin{cases}
   q^{(n)}_{n+k},  & |k|< N , \\
    0, & \text{otherwise}.
  \end{cases}
\end{equation}
Define a periodic function $\tilde v_t(\lambda)=\tilde v_t(\lambda+4+2d)$
 with $\tilde v^{(4)}_t\in L_2[\sigma_d]$, and such that
$\tilde v(\lambda)=V'(\lambda)$ for $ |\lambda|\le 2+d/2$.
Consider the standard Fourier  expansion for the function $\tilde v_t$
\begin{equation}\label{Four}
 \tilde v_t(\lambda)=\sum_{j=-\infty}^\infty v_{tj}e^{ij\kappa\lambda},\quad
  \kappa=\frac{\pi}{2+d},
\end{equation}
The first step in the proof of (\ref{lim_J}) is the lemma
\begin{lemma}\label{l:V'(J)}
If $V$ satisfies conditions C2-C3 and $V^{(4)}\in L_2[\sigma_d]$, then for any
$n^{1/3}<<N<n$ and any $|k|\le N/2$
 \begin{equation}\label{exp_v.1}\begin{array}{l}
\displaystyle V'_t({\mathcal{J}}^{(n)})_{n+k,n+k}=
\frac{t}{n}\varphi({\mathcal{J}}^{(0)})_{k,k}+\sum\mathcal{P}_{k-l}(t)\tilde q_l+\tilde
 r_k^{(0)}+O(||\tilde{\mathcal{J}}||/n)+O(N^{-7/2}),\\
\displaystyle V'_t({\mathcal{J}}^{(n)})_{n+k,n+k+1}=1-\tilde J_k+
\frac{t}{n}\varphi({\mathcal{J}}^{(0)})_{k,k+1}+\sum\mathcal{P}_{k-l}(t)\tilde J_{l}
+\tilde r_k^{(1)}\\
 \hfill +O(||\tilde{\mathcal{J}}||/n)+O(N^{-7/2}),
\end{array}\end{equation}
where for $\alpha=0,1$
 \begin{equation}\label{ti_r_k}
 \tilde r_k^{(\alpha)}=\sum_{j=-\infty}^{\infty} v_{tj}(ij\kappa )^2
 \displaystyle\int_0^1 ds_1\displaystyle\int_0^{1-s_1}ds_2
\left(e^{ij\kappa s_1{\mathcal{J}}^{(0)}}\tilde{\mathcal{J}}
e^{ij\kappa s_2{\mathcal{J}}^{(0)}}\tilde{\mathcal{J}}
 e^{ij\kappa (1-s_1-s_2)({\mathcal{J}}^{(0)}+\tilde{\mathcal{J}})}\right)_{k,k+\alpha}
\end{equation}
with $v_j$, $d$ defined in (\ref{Four}), and
\begin{equation}\label{P_l}
\quad \mathcal{P}_l(t)=\frac{1}{\pi}\int_{-\pi}^{\pi}
(P(2\cos(x/2))+t\tilde\varphi(2\cos(x/2))/n) e^{ilx}
dx,\end{equation}
with $P$ defined in (\ref{P}) and $\tilde\varphi$-some polynomial with coefficients
depending on $\varphi$.
\end{lemma}
\textbf{ Proof of Lemma \ref{l:V'(J)}}
By  Proposition 1 of \cite{S:07} it is enough to obtain (\ref{exp_v.1}) for
$\tilde v_t({\mathcal{J}}^{(0)}+\tilde{\mathcal{J}})_{n+k,n+k+\alpha}$. Using the
spectral theorem, we have
\[ \tilde v_t({\mathcal{J}}^{(0)}+\tilde{\mathcal{J}})_{k,k+\alpha}=
\sum_{j=-\infty}^{\infty}
\left(v_{tj} e^{ij\kappa ({\mathcal{J}}^{(0)}+\tilde{\mathcal{J}})}\right)_{k,k+\alpha}.\]
Applying the Duhamel formula  two times we get for $\alpha=0,1$
\begin{multline}\label{Four.1}
\tilde v_t({\mathcal{J}}^{(0)}+\tilde{\mathcal{J}})_{k,k+\alpha}=
\tilde v_t({\mathcal{J}}^{(0)})_{k,k+\alpha}\\+
\sum_{j=-\infty}^{\infty} v_{tj}(ij\kappa )
\int_0^1ds\left(e^{ij\kappa s{\mathcal{J}}^{(0)}}\tilde{\mathcal{J}}
e^{ij\kappa (1-s){\mathcal{J}}^{(0)}}\right)_{k,k+\alpha}+r_k^{(\alpha)}.\end{multline}
To find the the first term in (\ref{Four.1}) we use the relation, which follows
from coincidence $\tilde v(\lambda)=V'(\lambda)$, $\lambda\in [-2,2]$ and (\ref{si_eq})
\begin{multline}\label{int0}
\tilde v_t({\mathcal{J}}^{(0)})_{n+k,n+k+\alpha}= \frac{1}{2\pi}\int_{-\pi}^{\pi}
\tilde v_t(2\cos x)\cos^\alpha x\,dx\\=
\frac{1}{2\pi}\int_{-\pi}^{\pi} (V'(2\cos x)+t\varphi'(2\cos x)/n)\cos^\alpha x\,dx\\=
\frac{1}{\pi}\int_{-\pi}^{\pi}dx\int_{-2}^{2}\cos ^\alpha
x\frac{\rho(\lambda)d\lambda}{2\cos x-\lambda}+
\frac{t}{2\pi n}\int_{-\pi}^{\pi} \varphi'(2\cos x)/n)\cos^\alpha x\,dx=\alpha
+\frac{tc^{(\alpha)}}{n}.
\end{multline}
Besides,
since by the spectral theorem
\begin{equation}\label{exp(J_0)}
(e^{ij\kappa s{\mathcal{J}}^{(0)}})_{k,l}=\frac{1}{2\pi}\int_{-\pi}^{\pi}e^{ij\kappa s\cos
x} e^{i(k-l)x}dx =\mathbf{ J}_{k-l}(j\kappa s),
\end{equation}
where $\mathbf{ J}_{k}(s)$ is the Bessel function, and since $V'$ is an odd function,
we get for any $l$ and integer $\alpha$
\begin{multline*}\sum_{j=-\infty}^{\infty} v_{0j}(ij\kappa )
\int_0^1ds\left(e^{ij\kappa s{\mathcal{J}}^{(0)}}\right)_{k,l}
\left(e^{ij\kappa (1-s){\mathcal{J}}^{(0)}}\right)_{l\pm\alpha,k+1-\alpha}\\=
\frac{1}{(2\pi)^2}\int_{-\pi}^\pi\int_{-\pi}^\pi dx dy
\frac{V'(2\cos x)-V'(2\cos y)}{2\cos x-2\cos y}\cos\left((k-l)(x-y)+(\alpha(1\mp1)+1)y\right)
=0,
\end{multline*}
Hence, the linear terms with respect to $\tilde J_k$ in the first equation of
(\ref{exp_v.1}) and the linear terms with respect to $\tilde q_k$ in the second equation
give us only the contribution of the order $tn^{-1}||\tilde{\mathcal{J}}||$.
Besides, we derive from (\ref{Four.1}) that the operator $\P$ from
the second line of (\ref{exp_v.1}) can be represented in the form
\[
\mathcal{P}_{k-l}(t)=\delta_{k,l}+\int ds
\sum_{j=-\infty}^{\infty} v_{tj}(ij\kappa )
\left(e^{ij\kappa s{\mathcal{J}}^{(0)}}E^{(n+l)}
e^{ij\kappa (1-s){\mathcal{J}}^{(0)}}\right)_{k,k+1},
\]
where we denote by $E^{(l)}$ a matrix with entries:
\[E^{(l)}_{k,m}=\delta_{k,l}\delta_{m,l+1}+\delta_{k,l+1}\delta_{m,l}.\]
It is easy to see that $\mathcal{P}(t)$ is a Toeplitz matrix, so
its entries can be represented in the form
\[
 P_{l,k}(t)=P_{l-k}(t)=\frac{1}{2\pi}\int_{-\pi}^{\pi}e^{ilx}F(x,t)dx,
 \quad F(x,t)=\sum \mathcal{P}_l(t)e^{ilx}.
\]
Thus, we obtain
\begin{equation}\label{F^1}\begin{array}{rcl}
F(x,1)&=&1+\displaystyle\sum_{j}(ij\kappa )v_{tj}\int_0^1 ds_1\sum_l\dfrac{1}{4\pi^2}
\int_{-\pi}^{\pi}\int_{-\pi}^{\pi}e^{il(-x_1+x_2+x)}(1+e^{-i(x_1+x_2)})
\\
&&\cdot\exp\{2ij\kappa [s_1\cos x_1+(1-s_1)\cos
x_2]\}dx_1dx_2\\
&=&\displaystyle 1+\frac{1}{2\pi}\int_{-\pi}^{\pi} \dfrac{v_t(2\cos x_1)-v_t(2\cos(x_1-x))}
{\cos x_1-\cos(x_1-x)}(1+\cos (2x_1-x))dx_1\\
&=&\displaystyle 1+\frac{1}{2\pi}\int_{-\pi}^{\pi} v_t(2\cos x_1)\bigg(\frac{1+\cos (2x_1-x)}
{\cos x_1-\cos(x_1-x)}+\frac{1+\cos (2x_1+x)}
{\cos x_1-\cos(x_1+x)}\bigg)dx_1\\
&=&P(2\cos (x/2))+P(-2\cos (x/2))+t\tilde\varphi(2\cos (x/2))/n,
\end{array}\end{equation}
where in the last line we have used (\ref{int0}) and (\ref{P}). For the
linear operator in the first line of (\ref{exp_v.1}) the
calculations are similar.
Lemma \ref{l:V'(J)} is proved.$\square$

\medskip
Let us use (\ref{exp_v.1}) in (\ref{string}). We obtain for $k\le N/2$
\[\begin{array}{l}
\sum\mathcal{P}_{k-l}(t)\tilde q_{l}=-\dfrac{tc^{(0)}}{n}
-\tilde r_k^{(0)}+O(||\tilde{\mathcal{J}}||/n)+O(N^{-7/2}),\\
 \sum\mathcal{P}_{k-l}(t)\tilde J_{l}=\dfrac{k+1}{n}-\dfrac{tc^{(1)}}{n}+\tilde J_k^2
   -\tilde r_k^{(1)}+O(||\tilde{\mathcal{J}}||/n)+O(N^{-7/2}),
\end{array}\]
where $c^{(0)}$ and $c^{(1)}$ are defined in (\ref{int0}).
We would like to consider this system of equations like two linear equations in $l_2$.
To this end we set for $|k|> N/2$
\[\begin{array}{l}\tilde r_k^{(0)}=\sum\mathcal{P}_{k-l}(t)q_{l}, \\
\tilde r_k^{(1)}=\sum\mathcal{P}_{k-l}(t)\tilde J_{l}-\dfrac{k+1}{n}-\tilde J_k^2.
\end{array}\]
It follows from (\ref{P_l}) that the operator $\mathcal{P}$ has a bounded
inverse operator whose entries can be represented in the form
\begin{equation}\label{P^-1}
 (\mathcal{P}^{-1})_{k-l}=\frac{1}{4\pi}\int_{-\pi}^{\pi}
(P(2\cos(x/2))+t\tilde\varphi(2\cos(x/2))/n)^{-1} e^{i(k-l)x}
dx.
\end{equation}
Then
\begin{equation}\label{string.1}\begin{array}{l}
 q_{l}=-\sum\mathcal{P}^{-1}_{l-k}(0)\bigg(\dfrac{tc^{(0)}}{n}+O(||\tilde{\mathcal{J}}||/n)
   +\tilde r_k+O(N^{-7/2})\bigg),\\
\tilde J_{l}=\sum\mathcal{P}^{-1}_{l-k}(0)\bigg(\dfrac{k+1}{n}+\tilde J_k^2
-  \dfrac{tc^{(1)}}{n}+O(||\tilde{\mathcal{J}}||/n) -\tilde r_k+O(N^{-7/2})\bigg).
\end{array}\end{equation}
Moreover, since by assumption  $v'$ has fourth derivative from $L_2[-2,2]$, $P$
also does (see \cite{Mu:53}). Therefore, using  a
standard bound for the tails of the Fourier expansion of the function
$f$ with $f^{(p)}\in L_2[-\pi,\pi]$
\begin{equation}\label{tails_F}
\sum_{j>M}|f_k|\le M^{-p+1/2}\bigg(\sum|f_k|^2k^{2p}\bigg)^{1/2}\le CM^{-p+1/2} ,
\end{equation}
we have for any $M$
\begin{equation}\label{b_P^{-1}}
\sum_{|l|>M}|\mathcal{P}^{-1}_{l}|\le M^{-7/2},\quad
\sum_{|l|>M}|l||\mathcal{P}^{-1}_{l}|\le M^{-5/2},\quad
\sum_{|l|>M}|l|^2|\mathcal{P}^{-1}_{l}|\le M^{-3/2}.
\end{equation}
Besides, since $\mathcal{P}^{-1}_{l}=\mathcal{P}^{-1}_{-l}$, we have
\begin{equation}\label{Pk}
\sum_{l-k}\mathcal{P}^{-1}_{l-k}\frac{k+1}{n}=\frac{l+1}{n}\sum_{l-k}\mathcal{P}^{-1}_{l-k}
=\frac{1}{2P(2)}\,\frac{l+1}{n} .
\end{equation}
Using a trivial bound
\begin{equation}\label{triv_b}
 \left| \left(e^{ij\kappa s_1{\mathcal{J}}^{(0)}}\tilde{\mathcal{J}}
  e^{ij\kappa s_2{\mathcal{J}}^{(0)}}\tilde{\mathcal{J}}
 e^{ij\kappa (1-s_1-s_2)({\mathcal{J}}^{(0)}+\tilde{\mathcal{J}})}\right)_{k,k+1}
  \right|\le||\tilde{\mathcal{J}}||^2
\end{equation}
and (\ref{as_J}), we obtain first a rather crude bound
\begin{equation}\label{b_r.1}
|\tilde r_k^{(\alpha)}|\le C\left(|k|/{n}+n^{-1/2}\log^2n\right),\quad\alpha=0,1.
\end{equation}
This bound combined with (\ref{string.1}) and (\ref{tails_F}) give us
\begin{equation}\label{b_J}
|\tilde q_k|,|\tilde {\mathcal{J}}_k|\le C\left(|k|/{n}+
n^{-1/2}\log^2n+N^{-7/2}\right) .
\end{equation}
Now we use the bound, valid for any
 Jacobi matrix ${\mathcal{J}}$ with coefficients $%
J_{k,k+1}=J_{k+1,k}=a_{k}\in \mathbb{R}$, $|a_{k}|\leq A$.
Then there exist positive constants $C_0,\,C_{1},C_2$, depending
on $A$ such that  the
matrix elements of  $e^{it{\mathcal{J}}}$ satisfy the inequalities:
\begin{equation}
|(e^{it{\mathcal{J}}})_{k,j}|\leq C_0e^{-C_1|k-j|+C_2t}.
\label{exp_b}
\end{equation}
This bound follows from the representation
\[(e^{it{\mathcal{J}}})_{k,j}=-\frac{1}{2\pi i}\oint_le^{itz} R_{k,j}(z)dz,\]
where $ R=({\mathcal{J}}-z)^{-1}$, and from the Comb-Thomas type bound on the resolvent
of the Jacobi matrix (see \cite{RS})
\begin{equation}
|\mathcal{ R}_{k,j}(z)|\leq \frac{2}{|\Im z|}e^{-C_{1}'|\Im z|
|k-j|}+\frac{8}{|\Im z|^2}e^{-C_{1 }'|\Im z|(M-1)}.
\label{exp_b_R}
\end{equation}
Let us choose
\begin{equation}\label{cond_M}
M=\frac{C_1}{4C_2\kappa} n^{1/3},
\end{equation}
where $C_1$ and $C_2$ are the  constants from (\ref{exp_b}) and $\kappa=\pi(2+{\varepsilon })^{-1}$.
 Then (\ref{exp_b}) guarantee that for any $l,l':|l-l'|>n^{1/3}$ and any
$j:|j|<M$, $|t|\le 1$
\begin{equation}\label{exp_b1}
 |(e^{itdj{\mathcal{J}}^{(0)}})_{l,l'}|,
  |(e^{itdj({\mathcal{J}}^{(0)}+\tilde {\mathcal{J}}})_{l,l'}|
  \le Ce^{dC_2M-C_1|l-l'|}\le Ce^{-C_1n^{1/3}/3}e^{-C_1|l-l'|/3}.
\end{equation}
Now we split the sum in (\ref{ti_r_k}) in two parts $|j|< M$ and $|j|\ge M$.
\begin{multline}\label{ti_r_k.1}\tilde r_k^{(\alpha)}=\sum_{j=-\infty}^{\infty} v_j(ij\kappa )^2
 \displaystyle\sum_{l_1,l_2}\displaystyle\int_0^1 ds_1\displaystyle\int_0^{1-s_1}ds_2\\
\left(e^{ij\kappa s_1{\mathcal{J}}^{(0)}}\tilde{\mathcal{J}}\right)_{k,l_1}
\left(e^{ij\kappa s_2{\mathcal{J}}^{(0)}}\right)_{l_1,l_2}
\left(\tilde{\mathcal{J}} e^{ij\kappa (1-s_1-s_2)({\mathcal{J}}^{(0)}
+\tilde{\mathcal{J}})}\right)_{l_2,k+1}
=\sum_{|j|< M}+\sum_{|j|\ge M}.
\end{multline}
Then (\ref{exp_b1}) allows us to write
\begin{multline*}\sum_{|j|< M}=\sum_{|j|< M} v_j(ij\kappa )^2
 \displaystyle\sum_{l_1,l_2=k-[n^{1/3}]}^{k+[n^{1/3}]}
 \displaystyle\int_0^1 ds_1\displaystyle\int_0^{1-s_1}ds_2\\
\left(e^{ij\kappa s_1{\mathcal{J}}^{(0)}}\tilde{\mathcal{J}}\right)_{k,l_1}
\left(e^{ij\kappa s_2{\mathcal{J}}^{(0)}}\right)_{l_1,l_2}
\left(\tilde{\mathcal{J}} e^{ij\kappa (1-s_1-s_2)({\mathcal{J}}^{(0)}+
\tilde{\mathcal{J}})}\right)_{l_2,k+1}+O(e^{-Cn^{1/3}/3}).
\end{multline*}
Hence using (\ref{triv_b}) we obtain now
\begin{equation}\label{b_s.1}
\bigg|\sum_{|j|< M}\bigg|\le C\max_{l:|l-k-n|\le n^{1/3}}|\tilde J_l|^2.
\end{equation}
For $\sum_{|j|>M}$ we use (\ref{triv_b}) combined with (\ref{b_J}) and
(\ref{tails_F}) for the function $V'$. Then we get
\begin{equation}\label{b_s.2}
\bigg|\sum_{|j|\ge M}\bigg|\le C M^{-3/2}
\left(\left({N}/{n}\right)^2+n^{-1}\log^4n\right)\le Cn^{-1/2}\left({N}/{n}\right)^2
\end{equation}
and therefore
\begin{equation}\label{b_r.2}
|\tilde r_k^{(\alpha)}|\le C\left(\left((|k|+n^{1/3})/{n}\right)^2+n^{-1}\log^4n+N^{-7/2}+
n^{-1/2}\left({N}/{n}\right)^2\right) .
\end{equation}
 Using this bound in (\ref{string.1}) we obtain (\ref{lim_J}), but the bound for $r_k^{(\alpha)}$
 now has the form
\begin{equation}\label{b_r.3}|r_k^{(\alpha)}|\le C\left(\left(k/{n}\right)^2+n^{-1}\log^4n+N^{-7/2}+
n^{-1/2}\left({N}/{n}\right)^2\right)
\end{equation}
Now, using (\ref{lim_J}) with (\ref{b_r.3}) in (\ref{b_s.1}), and
setting $N=2[n^{1/2}]$ we obtain  the bound from (\ref{lim_J}) for $|k|\le n^{1/2}$. Then, setting
$N=2[n^{3/4}]$ and again using (\ref{lim_J}) with (\ref{b_r.3}) in (\ref{b_s.1}), we
obtain the bound from (\ref{lim_J}) for $n^{1/2}<k\le n^{3/4}$. And finally setting
$N=2[\tilde{\varepsilon } n]$, we obtain the bound from (\ref{lim_J})
 for $n^{3/4}<k\le \tilde{\varepsilon } n$.$\square$

\medskip

\textbf{Proof of Lemma \ref{l:A}}
The relation (\ref{lim_A}) is proved in \cite{S:07}. To prove
(\ref{b_A}) we need some extra definitions.
 We denote by $\mathcal{H}=l_2(-\infty,\infty)$ a Hilbert space of all
infinite sequences $\{x_i\}_{i=-\infty}^\infty$ with a standard
scalar product $(.,.)$ and a norm $||.||$. Let also $\{e_i\}_{i=-\infty}^\infty$ be
a standard basis in $\mathcal{H}$ and $I^{(-\infty,n)}$
be an orthogonal projection operator defined as
\begin{equation}\label{E}
I^{(-\infty,n)}e_i=\left\{\begin{array}{ll}e_i,& i<n,\\
0,& otherwise.\end{array}\right.
\end{equation}
For any infinite  matrix $\mathcal{ A}=\{A_{i,j}\}$
we will denote by
\begin{equation}\label{A_k,N}\begin{array}{l}
\mathcal{ A}^{(-\infty,n)}=I^{(-\infty,n)}\mathcal{A }I^{(-\infty,n)},\\
(\mathcal{ A}^{(-\infty,n)})^{-1}=
I^{(-\infty,n)}\bigg(I-I^{(-\infty,n)}+\mathcal{ A}^{(-\infty,n)}\bigg)^{-1}I^{(-\infty,n)},
\end{array}\end{equation}
so that $(\mathcal{A }^{(-\infty,n)})^{-1}$ is a block operator
which is inverse to $\mathcal{ A}^{(-\infty,n)}$ in the space
$I^{(-\infty,n)}\mathcal{H}$ and zero on the
$(I-I^{(-\infty,n)})\mathcal{H}$.

Besides, we will say that the matrix $\mathcal{ A}^{(-\infty,n)}$ \textit{is of the exponential
type},   if there exist constants $C$ and $c$, such that

\begin{equation}\label{e_pr}
 |A_{n-j,n-k}|\le Ce^{-c(|j|+|k|)}.
\end{equation}
Define infinite Toeplitz matrices $\mathcal{P}$ and  $\mathcal{V}^*$
by their entries
\begin{equation}\label{P,V}
    P_{j,k}=\frac{1}{2\pi }\int_{-\pi}^{\pi}e^{i(j-k)x}dxP(2\cos x),\quad
    V^{*}_{j,k}=\frac{\hbox{sign}(k-j)}{2\pi }\int_{-\pi}^{\pi}e^{i(j-k)x}dxV'(2\cos x),
\end{equation}
and let the entries $\mathcal{R}$ be defined in (\ref{R}). Then it is proved in \cite{S:07} that
for $|j|,|k|\le 2\log^2 n$
\begin{equation}\label{M^-1}
(\mathcal{M}^{(0,n)})^{-1}_{n-j,n-k}=
(\mathcal{R}^{(-\infty,n)})^{-1}\mathcal{D}^{(-\infty,n)})_{n-j,n-k}+b_{n-j}a_{n-k}+O(n^{-1/10}),
\end{equation}
where
\[a_k=((\mathcal{R}^{(,n)})^{-1}e_{n-1})_{k},\quad b_j=((\mathcal{R}^{(-\infty,n)})^{-1}r^*)_{j},\]
and the vector $r^*\in\mathcal{I}^{(0,n)}\mathcal{H}$ has components
$r^*_{n-i}=R_i$ ($i=2,4,\dots$) with $R_i$ defined by (\ref{R})
Let us prove that
\begin{equation}\label{F}
\mathcal{F}^{(-\infty,n)}:=(\mathcal{R}^{(-\infty,n)})^{-1}\mathcal{D}^{(-\infty,n)}-
\mathcal{V}^{*(-\infty,n)}
\end{equation}
is of the first type.
It is proved in \cite{S:07} (see Proposition 1) that
\begin{equation}\label{exp_b_1}\begin{array}{l}
|\mathcal{R}^{-1}_{n-j,n-k}|\le Ce^{-c|j-k|}\\
|(\mathcal{R}^{(-\infty,n)})^{-1}_{n-j,n-k}-\mathcal{R}^{-1}_{n-j,n-k}|\le
C\min\{e^{-c|j|};e^{-c|k|}\}\le Ce^{-c(|j|+|k|)/2} .
\end{array}\end{equation}
Hence,
\begin{multline*}
|\mathcal{F}^{(-\infty,n)}_{n-j,n-k}|\le \bigg|\sum_{l\ge 1}
\mathcal{P}_{n-j,n}\mathcal{D}_{n-l,n-k}-\mathcal{V}_{n-j,n-k}^*\bigg|+Ce^{-c|j|}
\sum_{l\ge 1}e^{-c|l|}e^{-c|l-k|}\\
\le\bigg|\sum_{l\ge 0}
\mathcal{P}_{n-j,n}\delta_{k,1}\bigg|+C'e^{-c(|j|+|k|)/2}
\le C_1e^{-c(|j|+|k|)/2}.
\end{multline*}
Besides, (\ref{exp_b_1}) imply
\begin{equation}\label{b_a,b}|a_k|\le Ce^{-c|k|},\quad |b_j|\le Ce^{-c|j|}.\end{equation}
It is easy to see that
\[-\frac{1}{2}\sum_k\mathcal{V}_{k,j}^{(n)}\epsilon\psi^{(n)}_{k}=
\frac{1}{n}(\epsilon\psi^{(n)}_{j})'=\frac{1}{n}\psi^{(n)}_{j},\]
where we denote $\mathcal{V}_{j,k}=\hbox{sign}(k-j)V'(J^{(n)})_{j,k}$,
and that for $j,k\ge 2\log^2n$
 \[(\mathcal{M}^{(-\infty,n)})^{-1}_{n-j,n-k}=\mathcal{V}_{n-j,n-k}+O(e^{-c\log^2n}).\]
Hence, if we denote
\[A_{j,k}^{(n)}=
(\mathcal{M}^{(-\infty,n)})^{-1}_{n-j,n-k}-\mathcal{V}_{n-j,n-k},
\quad
A_{j,k}=\mathcal{F}^{(0,n)})_{n-j,n-k}+b_{n-j}a_{n-k},\]
then $S_n$ is indeed represented in the form (\ref{S.1}),(\ref{lim_A}) is valid
because of  (\ref{lim_J}) and (\ref{M^-1}), and (\ref{b_A}) is valid
because we have proved that $\mathcal{F}^{(0,n)}$ is of the first type and
because of (\ref{b_a,b}).

$\square$

\end{document}